\documentclass{llncs}
\usepackage[utf8]{inputenc} 								
\usepackage[T1]{fontenc}    								
\usepackage{color}
\usepackage{hyperref}       								
\usepackage{url}            								
\usepackage{booktabs}       								
\usepackage{amsfonts}       								
\usepackage{nicefrac}       								
\usepackage{microtype}      								
\usepackage{graphicx} 										
\usepackage[labelfont=bf, labelsep=period]{caption}			
\usepackage{float} 											
\usepackage{makecell} 										
\usepackage{ragged2e} 										
\usepackage[dvipsnames]{xcolor} 							
\usepackage{soul} 											
\usepackage{algorithm}										
\usepackage{algpseudocode}									
\usepackage{amsmath}										
\usepackage{xr}												
\usepackage{letltxmacro} 									

\usepackage{cite}
\usepackage[noabbrev,nameinlink,capitalise]{cleveref}
\crefname{figure}{Fig.}{Figs.}
\Crefname{figure}{Figure}{Figures}
\crefname{table}{Table}{Tables}
\Crefname{table}{Table}{Tables}
\crefname{sec}{Section}{Sections}
\Crefname{sec}{Section}{Sections}
\newcommand{\creflabel}[2]{\hyperref[#1]{\cref*{#1}#2}}

\LetLtxMacro{\oldcref}{\cref}
\LetLtxMacro{\oldCref}{\Cref}

\hypersetup{
	pdftitle={Streaming Detector Data to NERSC for On-the-Fly Processing},
	pdfauthor={Samuel S. Welborn},
	pdfkeywords={streaming, high-performance computing, real-time processing},
}

\hypersetup{
	colorlinks=true,
	linkcolor=PineGreen,
	filecolor=magenta,
	urlcolor=blue,
	citecolor=black
}

\def\inst#1{$^{#1}$}
\def\orcidID#1{$^{[#1]}$} 

\def\email#1{{\tt#1}}

\begin{document}

\title{Accelerating Time-to-Science by Streaming Detector Data Directly into Perlmutter Compute Nodes}

\author{Samuel S. Welborn\inst{1,}\orcidID{0000-0002-7697-6347} \and
Chris Harris\inst{1,}\orcidID{0000-0002-1113-3728} \and
Peter Ercius\inst{1,}\orcidID{0000-0002-6762-9976} \and
Deborah J. Bard\inst{1,}\orcidID{0000-0002-5162-5153} \and
Bjoern Enders\inst{*}\inst{1,}\orcidID{0000-0002-6009-6281}
}

\authorrunning{S.S. Welborn et al.}
\titlerunning{Streaming Detector Data into Perlmutter Compute Nodes}
\institute{\inst{1}Lawrence Berkeley National Laboratory, Berkeley, CA 94720, USA \\
\email{swelborn@lbl.gov, cjh@lbl.gov, percius@lbl.gov, djbard@lbl.gov, \inst{*}benders@lbl.gov}}
\maketitle

\begin{abstract}
    Recent advancements in detector technology have significantly increased the size and complexity of experimental data, and high-performance computing (HPC) provides a path towards more efficient and timely data processing. However, movement of large data sets from acquisition systems to HPC centers introduces bottlenecks owing to storage I/O at both ends. This manuscript introduces a streaming workflow designed for an high data rate electron detector that streams data directly to compute node memory at the National Energy Research Scientific Computing Center (NERSC), thereby avoiding storage I/O. The new workflow deploys \textit{ZeroMQ}-based services for data production, aggregation, and distribution for on-the-fly processing, all coordinated through a distributed key-value store. The system is integrated with the detector's science gateway and utilizes the NERSC Superfacility API to initiate streaming jobs through a web-based frontend. Our approach achieves up to a 14-fold increase in data throughput and enhances predictability and reliability compared to a I/O-heavy file-based transfer workflow. Our work highlights the transformative potential of streaming workflows to expedite data analysis for time-sensitive experiments.

    \keywords{streaming \and 4D-STEM \and high-performance computing \and real-time processing}
\end{abstract}

\section{Introduction}

The transition from analog to digital data acquisitions and processing has greatly accelerated scientific discovery, but it also introduced the challenge of managing, processing, and interpreting an ever-expanding volume of data. In recent years, this challenge has intensified, with modern microscope detectors now achieving data generation rates five orders of magnitude greater than in the 1920s.\cite{Spurgeon2021-ym, rao2020deluge}

The National Energy Research Scientific Computing Center (NERSC) at Lawrence Berkeley National Laboratory (LBNL) responded to these challenges with the Superfacility Project.\cite{bard2022superfacility,enders2020cross} The project was designed to integrate experimental and observational science (EOS) facilities, many of which are incorporating high framerate detectors into their instruments, with state-of-the-art high-performance computing (HPC) resources. One of its notable achievements was a semi-automated file transfer and data reduction workflow developed for users of the National Center for Electron Microscopy (NCEM) facility of The Molecular Foundry (TMF), also at LBNL. Powered by NERSC's Superfacility API\cite{enders2020cross,bard2022superfacility} and the \textit{Distiller} web application,\cite{distiller} this workflow enables microscopists at NCEM to offload and process data from the 4D Camera\cite{ercius20234dcamera}—--an advanced detector that generates data at 480 Gb/s—--on NERSC compute nodes. Compared to processing at the edge on a single node, the NERSC workflow improved throughput by a factor of two.

Despite its impact, this workflow suffers from a large file I/O bottleneck. For example, a 695 GB dataset (\textasciitilde1 million detector frames) transferred by bbcp\cite{hanushevsky2002peer} from NCEM's local NFS buffer to NERSC scratch incurs delays of six or more minutes. These delays impact the microscopists' ability to make timely experimental decisions and impede real-time data analysis, highlighting the need for enhanced data management strategies that can support the high throughput demands of fast detectors.

This manuscript presents an approach to circumvent traditional file-based operations through data streaming. Utilizing \textit{ZeroMQ}, our new workflow facilitates direct data transfer from NCEM server RAM to NERSC compute node RAM for on-the-fly processing. This solution involves deploying several services to facilitate the transfer, including a data production service on the detector's data receiving servers, a data aggregation and fair-queuing distribution service at NCEM, and data consumption services at NERSC. We developed a \textit{ZeroMQ}-based distributed key-value store to connect and coordinate these services. Finally, to facilitate adoption of this new workflow, we extended enabled the creation of streaming sessions (compute jobs) from a web frontend.

\section{Background}
\label{sec:background}

TMF is a shared experimental facility that attracts researchers from many scientific disciplines to fabricate and analyze nanomaterials with state-of-the-art tools. The NCEM facility houses several advanced electron microscopes. Among these is the TEAM 0.5,\cite{ercius20234dcamera} a scanning transmission electron microscope (STEM) outfitted with the 4D Camera (\creflabel{fig:filetransfer}{b}) designed to rapidly capture large numbers of electron diffraction patterns. During data acquisition, a focused electron probe rasters across a sample in a 2D grid, pausing at each grid point for a predefined interval, the dwell time, to generate electron scattering events from the probe-sample interactions. The 4D Camera captures these events at 87 kHz on a 576 by 576 pixel array,\cite{ercius20234dcamera} resulting in a 4D dataset consisting of two sample $(x,y)$ and two detector coordinates$(q_x, q_y)$ leading to the name 4D-STEM. These complex datasets enable analytical methods like electron ptychography, which has gained traction in recent years for its ability to image atomic structure of a sample with high resolution.\cite{ophus20194dstem, Chen2021-dp}

\begin{figure}[H]
    \centering
    \includegraphics[width=\textwidth]{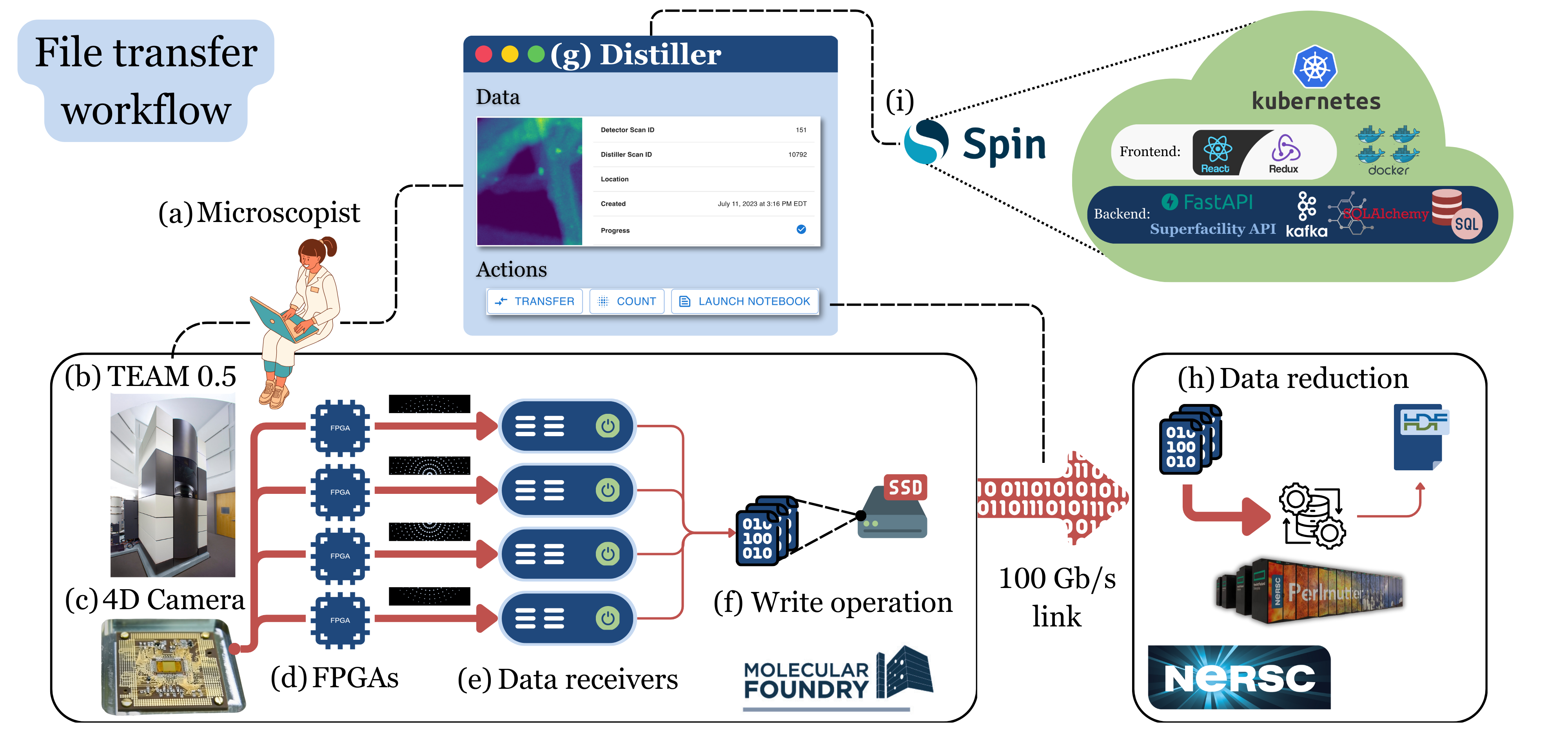}
    \caption{
        Schematic of the conventional file transfer workflow at The Molecular Foundry (TMF). A microscopist (a) takes an acquisition on the TEAM 0.5 microscope (b) with the 4D Camera (c). The data is read out from the detector by FPGAs (d) and sent upstream over UDP to data receiver servers (e) at 120 Gb/s per link. The receivers descramble the UDP packets and write files to a network file system (NFS) buffer. The microscopist can then interact with \textit{Distiller} (g) to transfer this data to NERSC for data reduction (h). \textit{Distiller} runs on Spin, NERSC's cloud-inspired infrastructure (i).
    }
    \label{fig:filetransfer}
\end{figure}

4D Camera acquisitions produce relatively large data volumes---approximately 695 GB for an acquisition of \textasciitilde1 million frames---in the same time as compared to traditional STEM imaging (megabytes per image). The data can be considered sparse, comprised of individual electron strike events, and can be compressed by an order of magnitude using a thresholding and peak finding algorithm (called "electron counting"). \textit{stempy}\cite{stempy} implements a highly parallelized version of this algorithm; however, the NCEM 10-core edge compute processing machine necessitates 10-12 minutes for a full dataset. During this time, new scans cannot be taken, thus imposing a 10-12 minute interval between acquisitions.

These wait times led to the development of a file transfer workflow (\creflabel{fig:filetransfer}) that offloads the processing onto NERSC compute nodes. In this workflow, a microscopist (\creflabel{fig:filetransfer}{a}) initiates an acquisition (\creflabel{fig:filetransfer}{b}) and the 4D Camera collects the scattered electrons (\creflabel{fig:filetransfer}{c}). During data capture, FPGAs (\creflabel{fig:filetransfer}{d}) facilitate readout from the detector, transmitting the data at 120 Gb/s per FPGA (480 Gb/s aggregate) to four data receiving servers (\creflabel{fig:filetransfer}{e}) utilizing the User Datagram Protocol (UDP). On these servers, a process is deployed that preallocates a significant fraction (\textasciitilde85\%) of the server's 256 GB RAM with an array of data structures. These structures contain both header information and a 144$\times$576 \texttt{uint16} array representing pixels from a single detector sector. Each detector frame is therefore initially dispersed across the four data receiving servers and can not be processed until they are recombined. The arrays are then flushed to multiple binary files on an 8 TB network file system (NFS) flash buffer (\creflabel{fig:filetransfer}{f}).

The microscopist tracks the write operation via the \textit{Distiller}\cite{distiller} web interface (\creflabel{fig:filetransfer}{g}), hosted on NERSC's \textit{Spin}\cite{spin} infrastructure (\creflabel{fig:filetransfer}{i}). \textit{Distiller}'s backend, leveraging \textit{FastAPI},\cite{lathkar2023high} \textit{Apache Kafka},\cite{garg2013apache} and a \textit{postgreSQL} database,\cite{drake2002practical} processes acquisition metadata in real time. Concurrently with data writing, a JSON file detailing the scan's ID and offload progress is used to update a \textit{FastAPI} database with a new scan record. Upon completion, users can launch data transfer and reduction jobs from the \textit{Distiller} frontend ("count" in \creflabel{fig:filetransfer}{g}). Orchestrated by \textit{FastAPI}, \textit{Kafka}, and an event-triggered job worker, this action will create a \textit{Slurm}\cite{yoo2003slurm} batch script using \textit{Jinja}\cite{nipkow2003jinja} templates and submit it to NERSC's realtime queue using the Superfacility API.\cite{enders2020cross} The job moves the raw files to NERSC scratch storage over a 100 Gb/s connection and sparsifies them according to the electron counting algorithm with \textit{stempy}\cite{stempy} (see also \cref{sec:consumers}).

While the workflow depicted in \cref{fig:filetransfer} effectively offloads data processing to NERSC and offers a user-friendly frontend for initiating data transfers, it incurs a notable performance cost due to four file I/O operations: initial writing to NFS at NCEM, reading and transferring data to NERSC, writing to NERSC's scratch system, and loading data from scratch into batch nodes. Our streaming workflow, detailed in the following section, completely bypasses this I/O bottleneck and significantly reduces the processing time.

\section{Methods}
\label{sec:methods}

Our streaming workflow extends the tooling discussed in \cref{sec:background} by integrating \textit{ZeroMQ} sockets over a wide-area-network (WAN) that facilitates data streaming from NCEM to NERSC. This setup employs two \textit{ZeroMQ} patterns: (1) the pipeline pattern, a work queue pattern where messages are fair-queued to downstream connections distributing messages evenly across workers, and (2) the clone pattern, which enables effective communication of system state through a distributed key-value store across the nodes. Additionally, we have extended \textit{Distiller} to provide NCEM users with access to the streaming workflow.

\subsection{Pipeline pattern}
\label{sec:pipeline}

\begin{figure}[H]
    \centering
    \includegraphics[width=\textwidth]{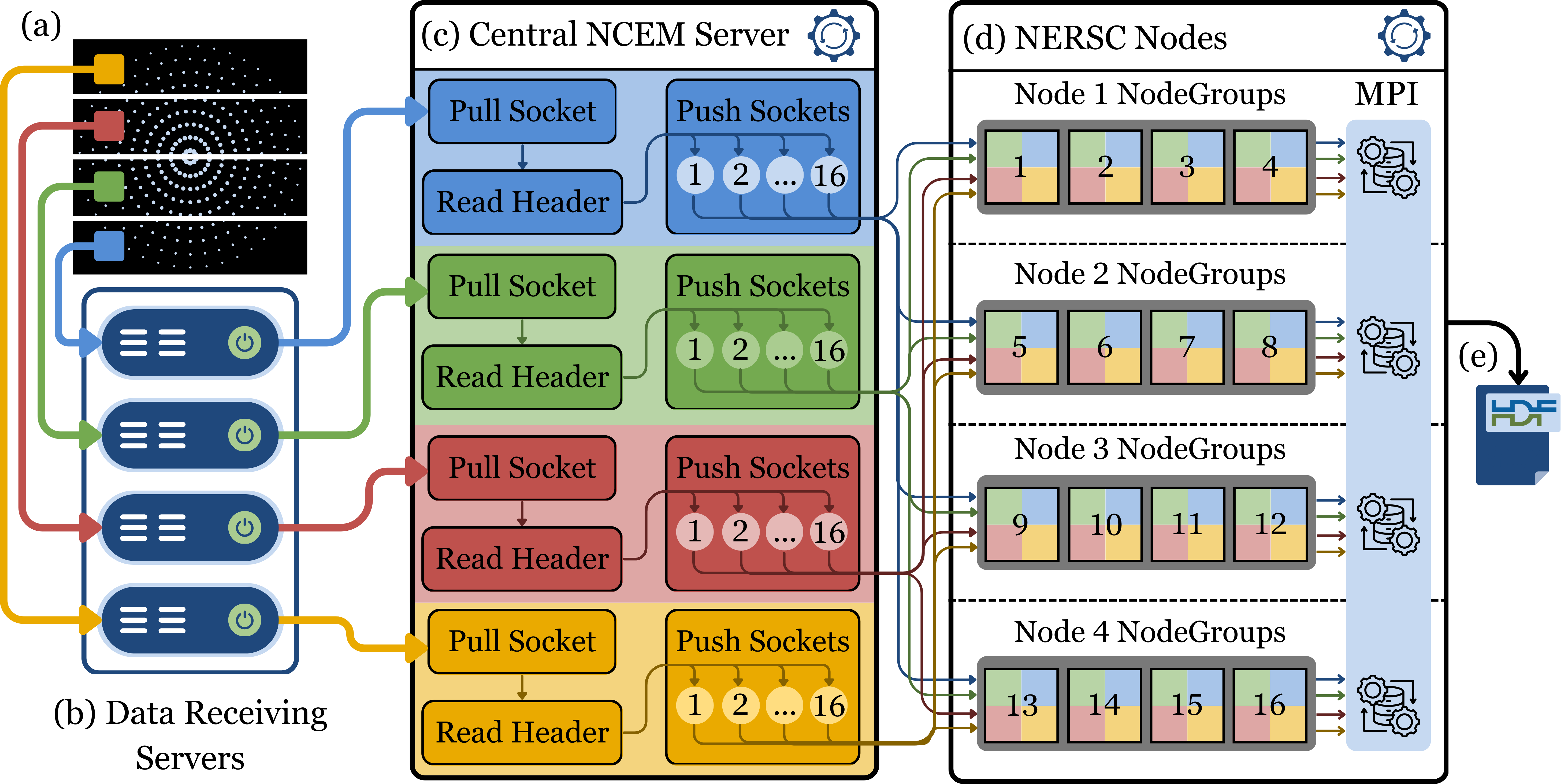}
    \caption{
        Schematic representation of the \textit{ZeroMQ} pipeline from NCEM to NERSC.
        (a) A 4D camera is partitioned into four 144$\times$576 sectors, each connected to a dedicated receiving server via FPGAs.
        (b) During data acquisition, the RAM of the data receiving servers is populated with sector data. The Producer objects on these servers push this data to a central aggregator service at NCEM.
        (c) Aggregators, denoted by varying colors, manage incoming messages by sequentially receiving them, extracting frame numbers from message headers, and forwarding the messages to the correct NodeGroup at NERSC.
        (d) On the compute nodes at NERSC, each node is subdivided into one or more NodeGroups (four per node depicted here). Each NodeGroup receives data from all NCEM Aggregators and forwards this data over the \texttt{inproc} protocol to stempy consumer threads. 
        (e) The data is processed and aggregated using \textit{stempy}'s electron counting methods with Message Passing Interface (MPI), consolidating the events in an HDF5 file.
    }
   \label{fig:pipeline_diagram}
\end{figure}

Described in Chapter 2 of the \textit{ZeroMQ} guide,\cite{zguide} the pipeline pattern fairly distributes messages from a push socket to all connected pull sockets. Microscope data is sensitive to dropped messages (i.e., data loss), and push sockets block instead of dropping messages when they reach their high water mark (HWM). This also ensures equitable data distribution across NERSC compute nodes. In our pipeline, the Data Receiving Servers use push sockets to send messages to an aggregator at NCEM (\creflabel{fig:network_diagram}{b-c}). The aggregator then relays these messages to the appropriate NERSC node (\creflabel{fig:network_diagram}{d}) for frame assembly and data reduction, \creflabel{fig:network_diagram}{e}. The pipeline includes two distinct messaging channels: the info channel informs downstream processes about the number of messages they can expect to receive and the data channel transmits the detector data. Color coding in \cref{fig:pipeline_diagram} signifies the origin and route of data from specific detector sectors to NERSC.

\subsubsection{Producers: Data Receiving Servers}
\label{sec:producers}

As mentioned in our description of the file transfer workflow (\cref{sec:background}), identical services running on each of the four data receiving servers ingest UDP packets from the detector (\creflabel{fig:network_diagram}{a-b}) and then flush data to disk. In our streaming workflow, the servers run similar application logic up until the data flushing stage. Each thread now uses push sockets to send data downstream to a central aggregator at NCEM (\creflabel{fig:network_diagram}{b-c}).

The threads first extract unique identifiers (UIDs) of the \texttt{NodeGroup}s (\cref{sec:consumers}) from the distributed key-value store (\cref{sec:clone}) and create a map of $\texttt{UID} \mapsto \texttt{n\_expected\_messages}$. For example, if a thread receives 100 sectors from the FPGA and ten \texttt{NodeGroup}s are available, it apportions ten sectors to each UID. This map, sent through the info channel, informs downstream processes of expected message volume.

The threads then continuously send two-part messages to the central server on the data channel. Each message is composed of a \textit{MsgPack}\cite{msgpack}-serialized header (part 1) and a 144$\times$576 \texttt{uint16} data array (part 2), representing a single frame sector. It is important to note that \textit{ZeroMQ} guarantees that all parts of a multi-part message are received, preventing message interleaving.

\subsubsection{Aggregator: Central NCEM Server}
\label{sec:routers}

The central aggregator server at NCEM runs four threads as depicted by the colored blocks in \creflabel{fig:pipeline_diagram}{c}. Each thread receives messages from all producer threads running on an individual Data Receiving Server (\creflabel{fig:pipeline_diagram}{b-c}) through the info channel. The threads then forward these messages to \texttt{NodeGroup}s (see \cref{sec:consumers}) based on each received sector's frame number. This approach ensures that the \texttt{Aggregator} threads divide the sector data evenly amongst the \texttt{NodeGroup}s, and that all four sectors of a single frame will end up on the same \texttt{NodeGroup}. Each thread executes the following procedure: First, it receives a $\texttt{UID} \mapsto \texttt{n\_expected\_messages}$ map for each connected producer thread and combines them. If a Data Receiving Server process has five threads each with $\texttt{UID} \mapsto \texttt{n\_expected\_messages}$, for example, we expect five maps to be received and the combined map to be $\texttt{UID} \mapsto \texttt{5*n\_expected\_messages}$. After combining, the thread pushes a message containing \texttt{n\_expected\_messages} to the appropriate downstream \texttt{NodeGroup} based on its \texttt{UID}. The thread then enters a tight pull-deserialize-push loop, illustrated in \creflabel{fig:pipeline_diagram}{c}. During each iteration, it receives two-part header/data message and deserializes the header to identify the sector's frame number. A push socket is selected based on the value of \texttt{frame\_number modulo n\_NodeGroups}, and the two-part message is forwarded on this socket. These push sockets are connected one-to-one to downstream \texttt{NodeGroup}s, which we have illustrated in \creflabel{fig:pipeline_diagram}{c-d}.

\subsubsection{Consumers: NERSC Nodes}
\label{sec:consumers}

At NERSC, \texttt{NodeGroup}s receive the messages routed to them by \texttt{Aggregator} threads. Each \texttt{NodeGroup} contains of four threads, as depicted by the four colored squares in \creflabel{fig:pipeline_diagram}{d}, that are connected one-to-one to an \texttt{Aggregator} thread. Each of these threads receives an info message to inform it of the expected message volume. Then, it enters a pull-push loop to receive header/data messages and send them over \texttt{inproc} to \textit{stempy} consumer threads.

We extended a \textit{stempy} \texttt{Reader} class, which normally reads from disk, to read from \textit{ZeroMQ} messages. As consumer threads pull two-part header/data messages from the \texttt{NodeGroup}s, the header is deserialized to extract the frame number and sector number, and data is stored in a map of $\texttt{frame number} \mapsto \texttt{sector number} \mapsto \texttt{data}$. Once the outer \texttt{frame number} map entry is populated with four sectors, the frame is complete and data reduction on that frame begins.

The electron counting algorithm employed for data reduction comprises several steps.\cite{Battaglia2009-cn} First, a subset of frames is chosen to establish thresholds for X-ray and background levels, utilizing binning techniques and Gaussian distribution fitting to the histogram generated from these samples. The Gaussian fit's initial parameters are derived from the sample mean and standard deviation. Specifically, the x-ray threshold is calculated as (\( \textit{mean} + M \times \textit{stddev} \)), where \(M=10\), while the background threshold is given by \( \textit{mean} + N \times \textit{stddev} \), where \( N \) is a tunable parameter (usually 4 or 4.5) set at runtime. After threshold determination, each frame undergoes a series of transformations. This includes subtracting a dark reference frame, if available, and applying the established X-ray and background thresholds. Following threshold application, local maxima are identified in relation to the nearest neighboring pixels; these maxima are interpreted as electron strike events.

As discussed in \cref{sec:producers}, data transmission in our pipeline begins with UDP-based communication from the FPGAs, a method that lacks guaranteed packet delivery. For very large scans, approximately 0.1\% of sectors are lost before the data enters our \textit{ZeroMQ} pipeline. To account for this, we only count complete frames until all expected messages are received and then count any incomplete frames. Eventually, all counted data is gathered on the first MPI rank. This aggregated dataset is then stored as a single HDF5 file on NERSC's scratch filesystem and asynchronously transferred to a long-term storage filesystem for later analysis.

\subsection{Clone}
\label{sec:clone}

\begin{figure}
    \centering
    \includegraphics[width=0.75\textwidth, keepaspectratio]{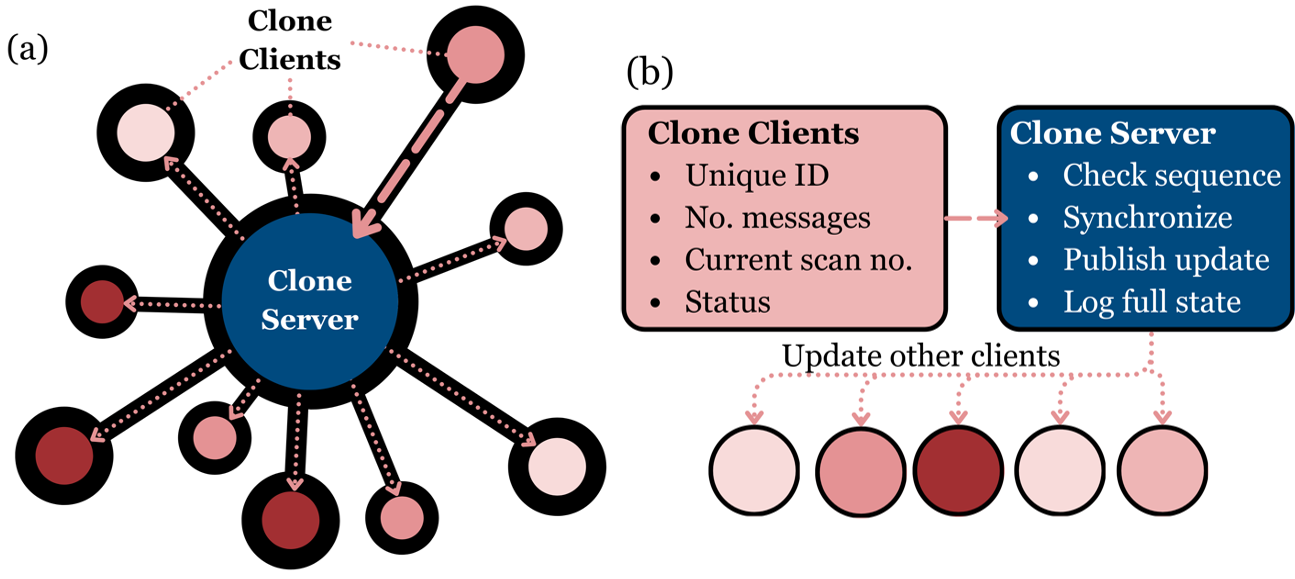}
    \caption{
        Network clients (producers, routers, and consumers) relay state updates through the central server, as schematized in (a).
        These updates include client-specific details like ID, sequence number, expected message count, scan number, and status (streaming, idle, etc.), as shown in (b).
        These updates are processed by the central server to adjust the network state and broadcast to all other clients.
    }
    \label{fig:network_diagram}
\end{figure}

Alongside the pipeline, we use the \href{https://zguide.zeromq.org/docs/chapter5/#Reliable-Pub-Sub-Clone-Pattern}{Clone} messaging pattern, a reliable publish-subscribe architecture detailed in Chapter 5 of the \textit{ZeroMQ} Guide.\cite{zguide} This pattern enables state synchronization across network nodes through a distributed key-value store. This pattern pushes the network state to a central server, which then disseminates these updates network-wide (\cref{fig:network_diagram}). Our adaptation introduces shared state objects to bridge client and pipeline threads, capturing and broadcasting essential details such as message count expectations, current scan numbers, and operational status (idle or streaming), as illustrated in \creflabel{fig:network_diagram}{b}. Further, we serialize the shared state objects with \textit{MsgPack}\cite{msgpack} for efficient transmission.

A key feature of the Clone pattern is dynamic network membership, enabling nodes to seamlessly join or exit. This flexibility is crucial for managing the lifecycle of streaming jobs at NERSC. Specifically, when we initiate a streaming job through \textit{Distiller}, the Producers and Aggregators are informed that new nodes have been added to the network, and consequently, they can begin to stream data to NERSC. When a job concludes, the Producers and Aggregators recognize that there are no available nodes, prompting a switch back to disk-based raw data storage.

This dynamic model serves several purposes: (1) It eliminates the necessity for an external notification system to inform Producers and Aggregators about node availability for streaming, thus removing the need for users to manually toggle between streaming and disk writing modes through a signaling mechanism. (2) It enables flexible node allocation for streaming jobs, ensuring the network can smoothly adjust to varying job sizes, whether they involve 2, 4, or more nodes. (3) It improves resiliency by defaulting to traditional disk writing as a reliable backup method when streaming nodes are not available.

\subsection{Initiating Consumers at NERSC through \textit{Distiller}}
\label{sec:workflow}

In developing the components in \cref{sec:pipeline} and \cref{sec:clone}, we recognized the necessity to accommodate end-users (the microsopists) who might not be familiar with high-performance computing (HPC) systems. Integrating the streaming workflow into an application familiar to the end-user is critical for adoption, so we upgraded \textit{Distiller} to include a streaming session manager. Initiating a session in many ways mirrors the user action event flow described in \cref{sec:background}, but now the event-triggered job worker can create \textit{Slurm} scripts that launch consumer services (\cref{sec:consumers}). Once available  (as detailed in Section \cref{sec:clone}), microscope acquisitions stream into Consumers. After an acquisition is transferred to long-term storage, \texttt{MPI rank 0} sends an asynchronous request to \textit{Distiller}'s FastAPI to update its location and session association in the database. The user is informed in \textit{Distiller}'s frontend when this acquisition is ready.

\section{Results}
\label{sec:results}

The streaming workflow demonstrates a faster and more consistent distribution of data transfer and processing times when compared to the file transfer approach based on our comparative analysis described below. This has two critical implications: (1) \textit{Acceleration.} The streaming pipeline significantly enhances data throughput, demonstrating approximately 14-fold and 5-fold increases for smaller and larger datasets, respectively. For larger datasets (i.e., 1024$\times$1024), data transfer and processing occur more quickly than the initial file write operation from RAM to disk in the traditional file transfer method --- the file-writing performance is approximately 4.6 GB/s, whereas the streaming pipeline achieves 7.2 GB/s. (2) \textit{Reliability.} The narrower time distribution indicates a more reliable and predictable system. This robustness is particularly advantageous for scheduling time-sensitive experiments and paves the way for future integration with automated systems. For example, the 695 GB streaming transfer has a standard deviation of $\pm$ 4.9 seconds ($\sigma_{s}$) compared to the file transfer method with $\pm$ 53.5 seconds ($\sigma_{ft}$).

\begin{figure}[H]
    \centering
    \includegraphics[width=0.9\textwidth]{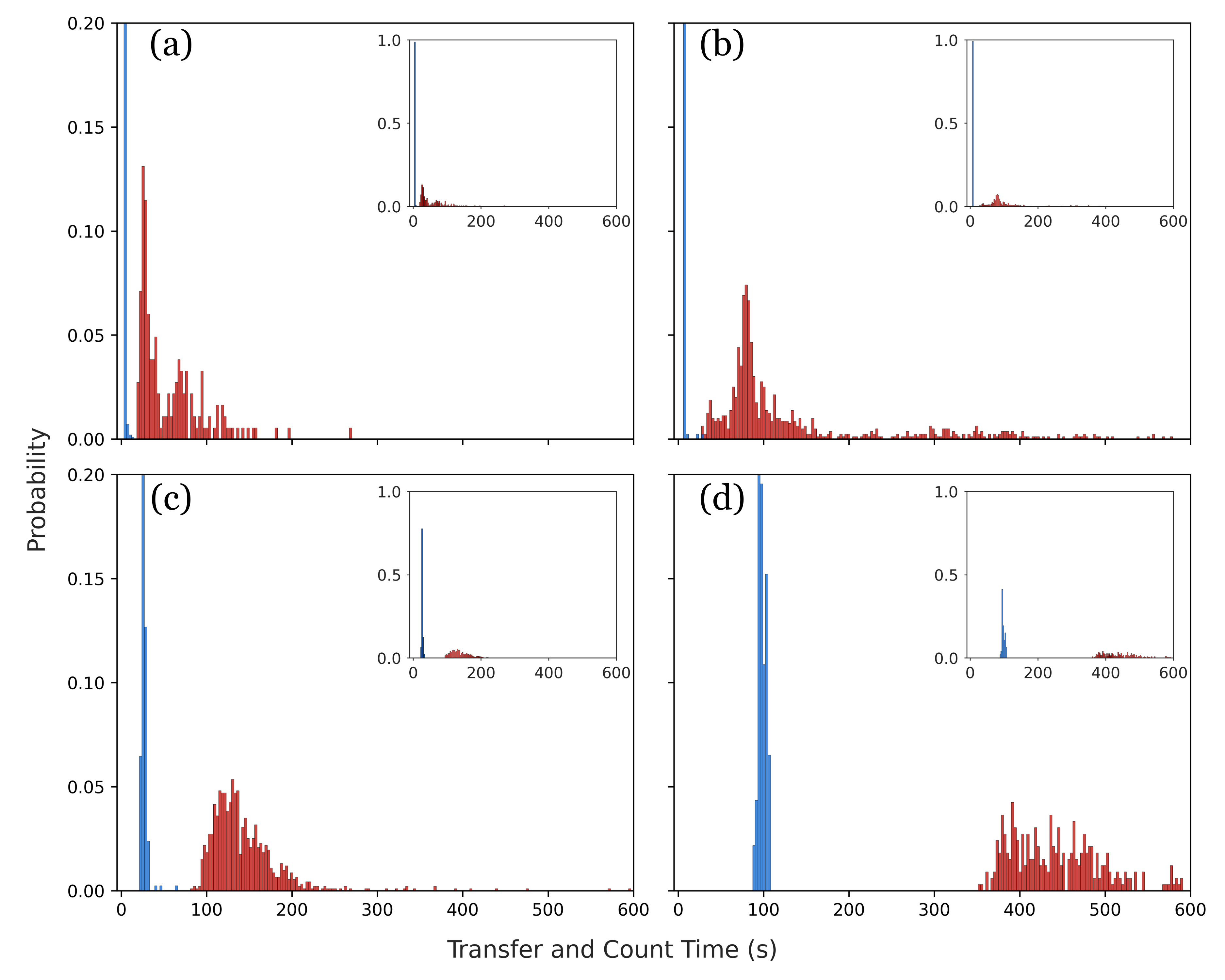}
    \caption{Histograms demonstrating superior performance of streaming (blue) compared with file transfer (red). (a-d) correspond to real space data dimensions 128$\times$128, 256$\times$256, 512$\times$512, and 1024$\times$1024, respectively. It is evident that the distribution of streaming times is both much narrower, and much faster than the distribution of file transfer times.}
\label{fig:transfer_histograms}
\end{figure}

We assessed the performance metrics for both pipelines on four standard real-space pixel dimensions (i.e., 2D array of probe positions) commonly used at NCEM: 128$\times$128 (10 GB), 256$\times$256 (43 GB), 512$\times$512 (173 GB), and 1024$\times$1024 (695 GB). To measure the streaming performance, we triggered the 4D Camera to send data at regular intervals while the electron beam was not active, so that the data collected did not contain electron events. Consequently, the metrics outlined here represent the optimal throughput for the current streaming architecture, as there is overhead on the NERSC consumer processes when events are present in the data. As we show in the accompanying repository, this overhead comes primarily from the only file I/O operation in the streaming pipeline --- the disk write operation of counted data at NERSC. This non-parallelized write operation occurs at around 340 MB/s. Both the sample area and pixel dimensions influence the number of events, and therefore the write time can be variable. For instance, for the sample used in the accompanying repository, the overhead for saving data from a 128$\times$128 acquisition is roughly 0.25 seconds, and extends to about 16 seconds for a 1024$\times$1024 acquisition covering the same sample area.

Metrics for the file transfer pipeline were sourced from the \textit{Distiller} database, which stores acquisition metadata and facilitates cross-referencing information with \textit{Slurm} such as queue time (the duration between job submission and job start times) and elapsed time (the time required for job execution). Aside from transfer and count times, another bottleneck in the file transfer workflow is the file write time (RAM to disk) at NCEM. To quantify this, we executed a series of `offload time' experiments for different common dataset sizes. 30 datasets were acquired for each configuration, and the average interval between initial file creation and final modification timestamps were calculated. These averages were added to the \textit{Slurm} elapsed times to encompass all steps of the file transfer pipeline. Finally, we take into account the overhead from writing the counted data at NERSC, discussed in the preceding paragraph. Since this overhead is variable, we conservatively subtract double the average write time for each data dimension (see accompanying repository for details) from the file transfer times. Metrics for the streaming pipeline were obtained through a similar analysis of timestamps, derived from NCEM logs and last modification times at NERSC.

\begin{table}
    \caption{Comparison of file transfer and streaming times for various data dimensions.}
    \centering
    \renewcommand{\arraystretch}{0.5}
    \begin{tabular}{ccccc}
    \toprule
    Data Dimension & Data Size (GB) & \makecell{File Transfer (s) \\ ($\mu_{ft} \pm \sigma_{ft}$)} & \makecell{Streaming (s) \\ ($\mu_{s} \pm \sigma_{s}$)} & \makecell{Enhancement \\ ($\mu_{ft}/\mu_{s}$)} \\
    \midrule
    128 x 128 x 576 x 576 & 10 GB & $52.0 \pm 30.6$ & $4.0 \pm 0.0$ & 13.0 \\
    256 x 256 x 576 x 576 & 43 GB & $92.3 \pm 38.6$ & $6.8 \pm 0.6$ & 13.6 \\
    512 x 512 x 576 x 576 & 173 GB & $138.5 \pm 28.2$ & $25.1 \pm 1.3$ & 5.5 \\
    1024 x 1024 x 576 x 576 & 695 GB & $442.6 \pm 53.5$ & $97.2 \pm 4.1$ & 4.6 \\
    \bottomrule
    \end{tabular}
    \label{tab:transfer_count_comparison}
    \end{table}

Outliers were identified and removed in accordance with standard practices before summarizing the data in \cref{tab:transfer_count_comparison}. Specifically, outliers were defined as observations that fall beyond \(1.5 \times \textit{IQR}\) (interquartile range), where \(\textit{IQR} = Q3 - Q1,\) and \(Q3\) and \(Q1\) represent the third and first quartiles, respectively. Such outliers usually indicate a file transfer workflow initiation failure. Data were then compiled into histograms and categorized by size for comparative analysis (\cref{fig:transfer_histograms}, \cref{tab:transfer_count_comparison}). 

Alongside this manuscript, we have incorporated a repository containing all data analyses to adhere to FAIR (Findable, Accessible, Interoperable, and Reusable) data principles.\cite{wilkinson2016fair}

\section{Related Work}
\label{sec:related}

Our literature review reveals multiple instances of software packages leveraging message queues and network data transfer within Data Acquisition (DAQ) systems. Common components across these systems typically include local RAM buffers for temporary data storage from detectors, a push/pull or publish/subscribe mechanism via sockets, and plugins for real-time data processing.

\begin{itemize}
    \item \textbf{PvaPy:} Recent efforts established a connection between the Advanced Photon Source (APS) and the Argonne Leadership Computing Facility (ALCF) using a streaming model derived from the Experimental Physics and Industrial Control System (EPICS).\cite{sinisa_streaming, dalesio1991epics} Utilizing \textit{PvaPy}\cite{veseli2015pvapy}, a Python interface for EPICS' pvAccess, a multi-producer, multi-consumer publish/subscribe network was constructed. This network achieved streaming rates exceeding 14 GB/s by employing multiple interconnected consumers. This work shows strong potential for integration with HPC centers, as it demonstrates high network throughput and uses a widely-adopted control system framework at beamlines across user facilities.
    \item \textbf{DUNE-DAQ:} The Deep Underground Neutrino Experiment (DUNE) generates neutrinos at Fermilab and detects them 800 miles away at Sanford Underground Research Facility to explore why the universe is made of matter. Their DAQ system employs \textit{ZeroMQ} wrappers and shared memory queues for bulk data transmission from detectors, event processing, and offloading to data writer processes.\cite{dunecollaboration2023dune}
    \item \textbf{ALFA:} Developed collaboratively by the Facility for Antiproton and Ion Research (FAIR) and A Large Ion Collider Experiment (ALICE) at CERN, the \textit{ALFA} framework utilizes \textit{FairMQ} for its transport layer.\cite{cern_alfa_original,cern_alfa} This layer consists of wrappers around \textit{ZeroMQ} sockets, called Devices, which are state machines that can be arranged in various topoligies to create communication channels. \textit{ALFA} also incorporates a processing layer with support for \textit{Apache Arrow} and \textit{ROOT}.\cite{alice_o2_software} 
    \item \textbf{ADARA:} The Accelerating Data Acquisition, Reduction, and Analysis (ADARA) system, built at Oak Ridge National Laboratory (ORNL), is another publish/subscribe system developed for real-time processing and visualization of Spallation Neutron Source (SNS) data.\cite{adara} The system uses a custom protocol on POSIX sockets to publish data, and supports both live and archived (persisted to disk) data streaming. The subscribers, which include the real-time visualization software Mantid and statistics/monitoring services, ingest the published data. While this system appears to be in use today,\cite{kilpatrick2023interactive} we cannot say for sure what the current state of the project is and if the custom protocol can handle the high data rates of modern detectors.
\end{itemize}

Although this list is not comprehensive, it underscores the diversity of existing solutions to similar challenges. Since these tools are designed with a specific DAQ system in mind (e.g., EPICS for PvaPy), using and retooling them for NCEM's DAQ would have posed a challenge. An ideal future tool would combine the best features of these systems and facilitate seamless integration with HPC centers with minimal application code changes.

\section{Conclusions and Outlook}
\label{sec:conclusions}

In response to the input/output (I/O) bottlenecks posed by conventional file transfer methods, this work introduced a \textit{ZeroMQ}-based pipeline to directly transfer large experimental datasets from the NCEM facility of TMF to compute nodes at NERSC. This approach effectively bypasses large disk storage operations at both ends, facilitating on-the-fly data processing.

Our results demonstrate a significant improvement in data throughput and system predictability, achieving up to a 14-fold increase in data transfer speed compared to NCEM's file transfer workflow. Further, we upgraded NCEM's user-facing web app, \textit{Distiller}, to enable microscopists to initiate and manage realtime jobs at NERSC from a web-based interface. These improvements reduce the turnaround time for microscopists and provides access to the new workflow without significant training overhead.

While our results confirm that streaming significantly reduces both processing delays and dependency on NERSC's shared file systems, the current implementation should be seen as a preliminary model. Its integration is tightly bound to NCEM-specific data formats and processing packages, indicating a necessity for a more universally adaptable tool that simplifies streaming to HPC facilities with minimal need for bespoke adjustments.

A more broadly applicable version of this tool should consider several key features, including but not limited to:

\begin{itemize}
    \item \textit{Semi-automated network management.} In our prototype implementation, the connection of various \textit{ZeroMQ} sockets is manually configured through specific IP addresses and port numbers within a configuration file. This procedure could be streamlined by leveraging a distributed key-value store as a dynamic IP address registry. Such an approach would automate the service discovery and connection process within the wide-area network (WAN), enabling new clients to register their IP addresses and identify connection partners.
    \item \textit{Decoupling of services from application code.} The current implementation also tightly binds services to NCEM-specific functionalities by subclassing the producers and consumers from our earlier file-transfer workflow. A future tool should aim for a decoupled architecture, where producer and consumer processes operate independently in separate memory spaces on the same machine. This separation would not only lower the technical threshold for adoption but also improve flexibility, allowing messages to be routed to several consumer processes for varying types of analysis.
\end{itemize}

\begin{credits}
    \subsubsection{\ackname} Work at the Molecular Foundry was supported by the Office of Science, Office of Basic Energy Sciences, of the U.S. Department of Energy under Contract No. DE-AC02-05CH11231. This research used resources of the National Energy Research Scientific Computing Center (NERSC), a U.S. Department of Energy Office of Science User Facility located at Lawrence Berkeley National Laboratory, operated under Contract No. DE-AC02-05CH11231 using NERSC awards BES-ERCAP0024753 and BES-ERCAP0024754. This work was partially funded by the US Department of Energy in the program "4D Camera Distillery: From Massive Electron Microscopy Scattering Data to Useful Information with AI/ML." We would like to thank Gatan, Inc. as well as P Denes, A Minor, J Ciston, C Ophus, J Joseph, Vamsi Vytla, and I Johnson who contributed to the development of the 4D Camera. 
    \subsubsection{\discintname} The authors have no competing interests to declare that are
    relevant to the content of this article. 
\end{credits}

\bibliography{references}

\end{document}